\documentclass[runningheads]{svmult}

\usepackage{makeidx}   % allows index generation
\usepackage{graphicx}  % standard LaTeX graphics tool
\usepackage{subeqnar}  % subnumbers individual equations
\usepackage{multicol}  % used for the two-column index
\usepackage{physprbb}  % modified textarea for proceedings,
\makeindex             % used for the subject index
\begin{document}
\title*{Spectroscopy of Planetary Nebulae in Sextans~A and Sextans~B\protect\newline }
\toctitle{Spectroscopy of Planetary Nebulae in Sextans~A and Sextans~B
\protect\newline }
% allows explicit linebreak for the table of content
%
%
\titlerunning{Planetary Nebulae in Sextans~A and Sextans~B}
% allows abbreviation of title, if the full title is too long
% to fit in the running head
%
\author{Laura Magrini\inst{1}
\and Pierre Leisy\inst{2,3}
\and Romano L. M. Corradi\inst{3}
\and Mario Perinotto\inst{1}
\and Antonio Mampaso\inst{2}
\and Jos\'{e} V\'{\i}lchez\inst{4}
}
\authorrunning{Magrini et al.}
% if there are more than two authors,
% please abbreviate author list for running head
%
%
\institute{Dipartimento di Astronomia e Scienza dello Spazio, Universit\'a di 
Firenze, Italy 
\and Instituto de Astrof\'{\i}sica de Canarias, Tenerife, Canarias, Spain 
\and Isaac Newton Group of Telescopes, La Palma, Canarias, Spain
\and Instituto de Astrof\'{\i}sica de Andaluc\'{\i}a, Granada, Spain}
\maketitle              % typesets the title of the contribution

\begin{abstract}
Sextans~A and Sextans~B are two dIr galaxies situated in the outskirts of
the Local Group (LG), in which both PNe and H~II regions have been detected
(Jacoby \& Lesser \cite{jl81}, Magrini et al. \cite{m02}, \cite{m03}). 
We present spectroscopic observations of PNe and HII regions 
in these two galaxies (obtained with  the VLT)
Preliminary results about PNe physico-chemical properties  are presented.   

\end{abstract}

\section{The galaxies: Sextans~A and Sextans~B }
Sextans~A and Sextans~B are both dwarf Irregular galaxies (Ir V and Ir
IV-V morphological types, respectively, cf. van den Bergh
\cite{bergh}, hereafter vdB00) with approximately the same V luminosity, 
belonging to the outer fringes of the
Local Group.  Their distances from the barycenter of the LG are quite
similar (1.60 and 1.72 Mpc, see vdB00) and their separation on the sky
is relatively small ($\sim$ 10 degree), which corresponds to about
280~kpc.  Moreover their velocity difference is only 23 $\pm$6 km
s$^{-1}$.  These results suggest the hypothesis of a common formation
of these two galaxies, probably together with NGC~3109 and the
Antlia galaxy.  Considering the mean distance of the four galaxies
from the barycenter of the LG, 1.7~Mpc, this sub-group is located
beyond the zero velocity surface of the LG (cf. vdB00) and thus it can
be considered the nearest external group of galaxies.
As said by Mateo (\cite{mateo}) ``No two Local Group dwarfs have the same star-formation 
history''. This is true also when comparing very similar galaxies.  
In fact, in spite of a probable common formation, Sextans~A and Sextans~B show
a different star formation history as indicated by the different
amounts of stars in the various evolutionary phases, and also
reflected in the different number of PNe and HII regions observed.
Sextans A shows a large amount of old stellar population, whereas 
it has a modest intermediate (4-10~Gyrs ago) and recent star formation (1-4~Gyrs ago). 
On the other hand, Sextans B exhibits a very strong recent star 
formation (1-4~Gyrs ago), together with a very old population (cf. Mateo's (\cite{mateo}
review about the star formation history of the LG dwarfs). 

In this context, PNe  represent an useful tool to trace the chemical enrichment history 
of galaxies, moreover when compared with chemical abundance derived from H~II regions, because they 
are tracers of stellar population in a wide range of ages, from old to intermediate age.

\section {Observations and data reduction}
\begin{figure}[h!]
\begin{center}
\includegraphics[width=0.7\textwidth, angle=270]{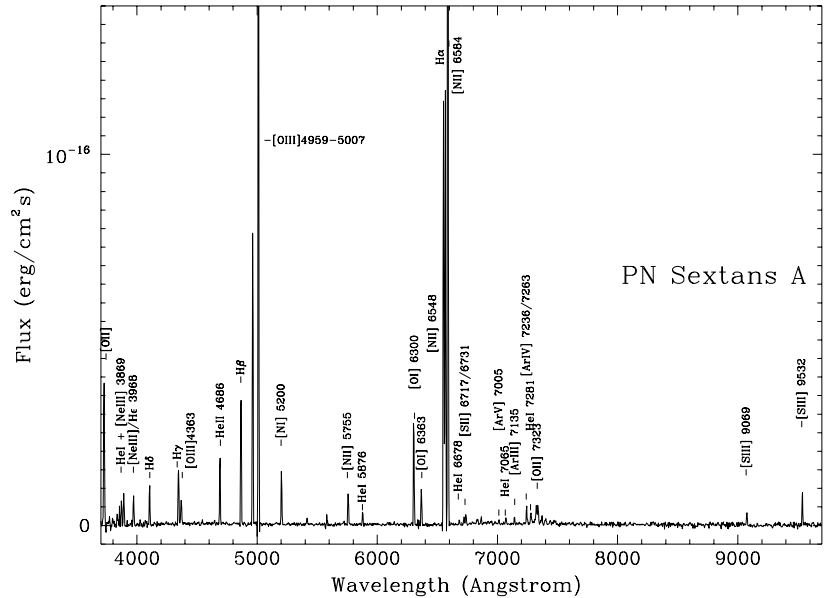}
\end{center}
\caption[]{VLT spectra of the planetary nebula in Sextans A}
\label{fig1}
\end{figure}

\begin{figure}[h!]
\begin{center}
\includegraphics[width=0.6\textwidth, angle=270]{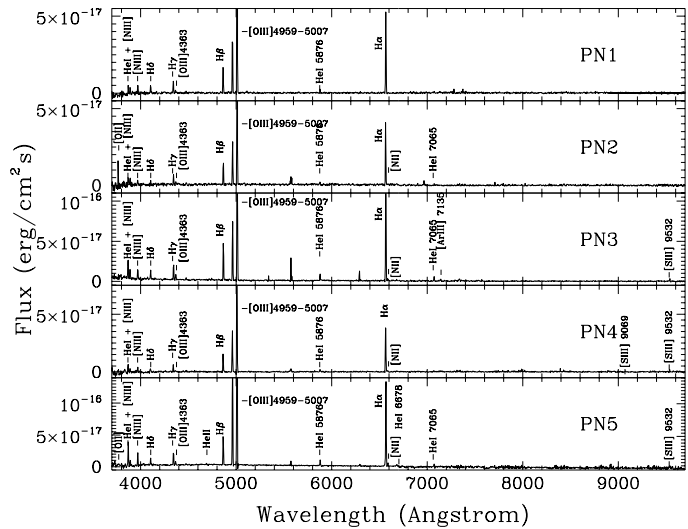}
\end{center}
\caption[]{VLT spectra of the five planetary nebulae in Sextans B. Identification numbers 
come from in Magrini et al. (\cite{m02})}
\label{fig2}
\end{figure}

PNe (1 in Sextans~A and 5 in Sextans~B) and H~II regions (10 and 9, respectively)  
have been observed in December 2003 with VLT (ESO, Paranal) equipped with
FORS2 spectrograph in multi-object spectroscopy (MOS) observing mode.  
The FORS2 spectrograph was used with two setup: the
300~V and 300~I gratings  providing a
dispersion of 3.0 \AA/pixel.  The total resulting spectral range, from
$\sim$3200\AA\, to $\sim$10000 \AA, includes basic lines needed for
the determination of chemical abundances.  For each galaxy, a total of
six exposures were taken, three with the 300~V grating (total exposure
time 5400~s) and three using the 300~I one (3600~s).  A
spectrophotometric standard star, GD~108, was observed once during each night.  
The data were reduced using the IRAF LONGSLIT package and
complemented with MIDAS (MOS and LONG packages).  
The spectra are presented in
Figs. \ref{fig1} and \ref{fig2}. Identified emission lines are marked
with the name of the corresponding ion and/or wavelength.
 
\section{Chemical abundances of planetary nebulae}

Chemical abundances have been derived with the classic Ionization
Correction Factors (ICFs) method, following Kingsburgh \& Barlow
(\cite{kb94}).  As the [O~III] 4363  emission line was always measurable 
with a sufficiently high signal to noise ratio, the 
[O~III] electron temperature was derived for all
PNe.  
As the nitrogen emission lines are so intense in Sextans A, 
the  [N~II] 5755 is observed and an electron temperature  is 
derived. 
[S~III] infrared emission lines were detected in 4
PNe. They allow a better accuracy in the measurement of sulphur
abundance, especially in low metallicity environments where [S~II] lines
are extremely faint.

In addition, PNe have been modelled with the phoionization code CLOUDY
94.00 (Ferland et al. \cite{ferland}), assuming a blackbody central
star with effective temperature derived using the Ambartsumian's
(\cite{ambar}) or Gurzadyan's (\cite{gurzadyan}) methods, and a
spherical nebula with constant density, derived from [S~II] 6717/6731
\AA\, flux ratio when available or assumed equal to 3000 cm$^{-3}$ in the other cases. 
For further details on the modelling procedure see Magrini et
al. (\cite{M04}) and Perinotto et al. (\cite{P04}). 
The chemical abundances derived with the two methods are 
in good agreement one to each other. The CLOUDY model allowed to derive also 
some nebular and stellar parameters as the radius of the 
nebula, the stellar luminosity and temperature (see Table~\ref{Tab2}).
The preliminary  chemical abundances and their errors, derived with the ICFs method, 
are shown in Table~\ref{Tab1}.
\begin{table}
\caption{Chemical abundances of Sextans~A and Sextans~B planetary nebulae derived with the ICFs method.
Metal abundances are expressed in 12+$\log$(X/H). }
\begin{center}
{\scriptsize
\renewcommand{\arraystretch}{1.4}
\setlength\tabcolsep{5pt}
\begin{tabular}{lllllll}
\hline\noalign{\smallskip}
Name & He/H & N/H & O/H & Ne/N & Ar/H & S/H \\
\noalign{\smallskip}
\hline
\noalign{\smallskip}
SexA-PN	 & 0.085$\pm$0.01 & 8.45$\pm$0.1& 8.0$\pm$0.1 & 6.7$\pm$0.2 & 5.2$\pm$0.3 & 5.8$\pm$0.1\\ 
SexB-PN1 & 0.082$\pm$0.02 & -		 & 8.0$\pm$0.2 & 7.0$\pm$0.3 & -	           & -		  \\ 
SexB-PN2 & 0.05$\pm$0.03 & -	 	 & 7.65$\pm$0.3& 6.7$\pm$0.3 & -		   & -		  \\ 	
SexB-PN3 & 0.083$\pm$0.01 & 5.4$\dag$   & 7.5$\pm$0.2 & 6.9$\pm$0.3 & 5.5$\pm$0.3 & -\\ 
SexB-PN4 & 0.084$\pm$0.02 & 5.5$\dag$	 & 8.0$\pm$0.3 & 7.0$\pm$0.3 & - 		   & 5.4$\ddag$ \\ 
SexB-PN5 & 0.118$\pm$0.02 & 8.0$\pm$0.3 & 8.2$\pm$0.1 & 7.65$\pm$0.2 & - 		   & -\\ 
\hline
\end{tabular}
}
\end{center}
\label{Tab1}
Note: $\dag$ this is the ionic N$^{+}$ abundance. 
$\ddag$ this is the ionic S$^{2+}$ abundance.  In both $\dag$ and $\ddag$ cases, ICF(N) and  ICF(S), respectively,  
were  not computed because [O~II] lines were not detected.
\end{table}

\begin{table}
\caption{Some nebular and stellar parameters of Sextans~A and Sextans~B planetary nebulae.
$\dag$ Derived with the CLOUDY model.  T$_{\star}$ marked with : are upper limits 
to the stellar effective temperature obtained when the He~II 4686 \AA\ was not
measurable. }
\begin{center}
{\scriptsize
\renewcommand{\arraystretch}{1.4}
\setlength\tabcolsep{5pt}
\begin{tabular}{lllllccl}
\hline\noalign{\smallskip}
Name 	 & N$_e$ & T$_e$$_{\rm{[O~III]}}$ & T$_e$$_{\rm{[N~II]}}$ &Radius(pc)$\dag$ & T$_{\star}$$\dag$ & log(L$_{\star}$/L$_{\odot}$)$\dag$ & c$\beta$\\
\noalign{\smallskip}
\hline
\noalign{\smallskip}
SexA-PN	 & 2700 & 21000 		& 13400		      &0.13 	  & 190000      & 3.8 & 0.22\\
SexB-PN1 & -	& 12900			& -		      &0.08	  & 50000:	& 3.2 & 0.11\\	
SexB-PN2 & -	& 17300			& -		      &0.02	  & 73000:	& 3.3 & -   \\ 
SexB-PN3 & -	& 17800			& -		      &0.03	  & 74000:	& 3.5 & 0.11\\
SexB-PN4 & -	& 13300			& -		      &0.03	  & 70000:      & 3.1 & -   \\
SexB-PN5 & -	& 12700			& -		      &0.01	  & 84000	& 3.4  & 0.14\\
\hline
\end{tabular}
}
\end{center}
\label{Tab2}
\end{table}

Although the definition of Type I PNe is a function of metallicity of the host galaxy (cf. Magrini et al. \cite{M04}), 
as a first approximation, we applied the Galactic Type~I definition by 
Kingsburgh \& Barlow (\cite{kb94}),  $\log$(N/O)$>$-0.1. 
With this definition, the Sextans~A PN is a Type~I PN, while the Sextans~B PNe are non-Type~I. 
Using the stellar luminosity and temperature derived with the CLOUDY model,  
we compared the location of the six PNe in the $\log$T$_{\star}$-$\log$L$_{\star}$ space 
with the H-burning evolutionary tracks from Vassiliadis \& Wood (\cite{vass}).
We found a more massive progenitor for the Sextans~A PN ($\sim$2.5M$_{\odot}$) than 
those of Sextans~B PNe ($\sim$1-1.5M$_{\odot}$), in agreement with its Type~I nature.

Another interesting aspect from Table~\ref{Tab1} is the large range of O/H, $\sim$0.7~dex,  
of Sextans~B PNe in spite of their similar evolutionary age. 
Since oxygen is not processed by PN progenitors, this represents 
a sign of the dishomogeneity of the interstellar medium at the time 
of the formation of the PN progenitors.
We plan to compare this dishomogeneity with the spatial distribution 
of H~II regions chemical abundances.

\section{The metallicity-luminosity relationship using PNe}
 \begin{figure}[h!]
\begin{center}
\includegraphics[width=0.5\textwidth, angle=270]{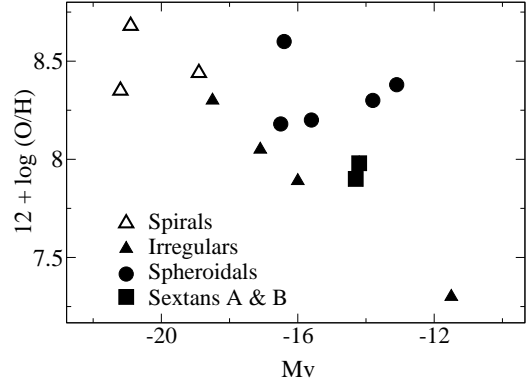}
\end{center}
\caption[]{Relationship of  V-luminosity vs. oxygen abundance 
from PNe in the LG. PN abundances: Sextans A and Sextans B from this work, 
other galaxies data present in literature.}
\label{fig3}
\end{figure}
In Fig.~\ref{fig3}, the relationship of  V-luminosity vs. oxygen abundance 
from PNe in the LG is shown. Galaxies are identified with different symbols 
according to their morphological type.    
A different behaviour of star forming  (irregulars and spirals) 
and non-star forming  (spheroidals) galaxies was already noted by several authors (cf. Mateo \cite{mateo}): spheroidal 
galaxies are generally metal richer than irregular galaxies with the same V luminosity. 
There are several hypotesys about this  behaviour (cf. Pilyugin \cite{pil}), but a satisfatctory 
understanding is still missing. 
The uniform way to derive chemical abundances with PNe, which are present in galaxies 
of every morphological type, will help to understand the origins of this  fundamental relation.

%INDEX%%%%%%%%%%%%%%%%%%%%%%%%%%%%%%%%%%%%%%%%%%%%%%%%%%%%%%%%%%%%%%%
% Please check with the editor of your book whether he plans to
% include a "mutual" subject index - if so, please code your entries
% in the standard syntax. For your own purposes you may print your
% "personal" index by using the following commands:
%
%\clearpage
%\addcontentsline{toc}{section}{Index}
%\flushbottom
%\printindex
%%%%%%%%%%%%%%%%%%%%%%%%%%%%%%%%%%%%%%%%%%%%%%%%%%%%%%%%%%%%%%%%%%%%%

\end{document}